\documentclass[english]{article}
\usepackage[T1]{fontenc}
\usepackage[latin9]{inputenc}
\usepackage{textcomp}
\usepackage{amsmath}
\usepackage{graphicx}

\makeatletter
\newcommand{\lyxaddress}[1]{
\par {\raggedright #1
\vspace{1.4em}
\noindent\par}
}

\let\jnl@style=\rm
\def\ref@jnl#1{{\jnl@style#1}}

\def\aj{\ref@jnl{AJ}}                   
\def\actaa{\ref@jnl{Acta Astron.}}      
\def\araa{\ref@jnl{ARA\&A}}             
\def\apj{\ref@jnl{ApJ}}                 
\def\apjl{\ref@jnl{ApJ}}                
\def\apjs{\ref@jnl{ApJS}}               
\def\ao{\ref@jnl{Appl.~Opt.}}           
\def\apss{\ref@jnl{Ap\&SS}}             
\def\aap{\ref@jnl{A\&A}}                
\def\aapr{\ref@jnl{A\&A~Rev.}}          
\def\aaps{\ref@jnl{A\&AS}}              
\def\azh{\ref@jnl{AZh}}                 
\def\baas{\ref@jnl{BAAS}}               
\def\bac{\ref@jnl{Bull. astr. Inst. Czechosl.}}
\def\caa{\ref@jnl{Chinese Astron. Astrophys.}}
\def\cjaa{\ref@jnl{Chinese J. Astron. Astrophys.}}
\def\icarus{\ref@jnl{Icarus}}           
\def\jcap{\ref@jnl{J. Cosmology Astropart. Phys.}}
\def\jrasc{\ref@jnl{JRASC}}             
\def\memras{\ref@jnl{MmRAS}}            
\def\mnras{\ref@jnl{MNRAS}}             
\def\na{\ref@jnl{New A}}                
\def\nar{\ref@jnl{New A Rev.}}          
\def\pra{\ref@jnl{Phys.~Rev.~A}}        
\def\prb{\ref@jnl{Phys.~Rev.~B}}        
\def\prc{\ref@jnl{Phys.~Rev.~C}}        
\def\prd{\ref@jnl{Phys.~Rev.~D}}        
\def\pre{\ref@jnl{Phys.~Rev.~E}}        
\def\prl{\ref@jnl{Phys.~Rev.~Lett.}}    
\def\pasa{\ref@jnl{PASA}}               
\def\pasp{\ref@jnl{PASP}}               
\def\pasj{\ref@jnl{PASJ}}               
\def\rmxaa{\ref@jnl{Rev. Mexicana Astron. Astrofis.}}%
\def\qjras{\ref@jnl{QJRAS}}             
\def\skytel{\ref@jnl{S\&T}}             
\def\solphys{\ref@jnl{Sol.~Phys.}}      
\def\sovast{\ref@jnl{Soviet~Ast.}}      
\def\ssr{\ref@jnl{Space~Sci.~Rev.}}     
\def\zap{\ref@jnl{ZAp}}                 
\def\nat{\ref@jnl{Nature}}              
\def\iaucirc{\ref@jnl{IAU~Circ.}}       
\def\aplett{\ref@jnl{Astrophys.~Lett.}} 
\def\apspr{\ref@jnl{Astrophys.~Space~Phys.~Res.}}
\def\bain{\ref@jnl{Bull.~Astron.~Inst.~Netherlands}} 
\def\fcp{\ref@jnl{Fund.~Cosmic~Phys.}}  
\def\gca{\ref@jnl{Geochim.~Cosmochim.~Acta}}   
\def\grl{\ref@jnl{Geophys.~Res.~Lett.}} 
\def\jcp{\ref@jnl{J.~Chem.~Phys.}}      
\def\jgr{\ref@jnl{J.~Geophys.~Res.}}    
\def\jqsrt{\ref@jnl{J.~Quant.~Spec.~Radiat.~Transf.}}
\def\memsai{\ref@jnl{Mem.~Soc.~Astron.~Italiana}}
\def\nphysa{\ref@jnl{Nucl.~Phys.~A}}   
\def\physrep{\ref@jnl{Phys.~Rep.}}   
\def\physscr{\ref@jnl{Phys.~Scr}}   
\def\planss{\ref@jnl{Planet.~Space~Sci.}}   
\def\procspie{\ref@jnl{Proc.~SPIE}}   

\makeatother

\usepackage{babel}
\begin{document}

\title{Angular Density Perturbations to Filled Type I Strong Explosions}

\author{Almog Yalinewich$^{1}$ and Re'em Sari$^{1,2}$}

\maketitle

\lyxaddress{$^{1}$Racah Institute of Physics, the Hebrew University, 91904,
Jerusalem, Israel\\
$^{2}$California Institute of Technology, MC 130-33, Pasadena, CA
91125}
\begin{abstract}
\noindent In this paper we extend the Sedov - Taylor - Von Neumann
model for a strong explosion to account for small angular and radial
variations in the density. We assume that the density profile is given
by $\rho\left(r,\theta,\phi\right)=kr^{-\omega}\left(1+\varepsilon\left(\frac{r}{r_{0}}\right)^{q}Y_{lm}\left(\theta,\phi\right)\right)$,
where $\varepsilon\ll1$ and $\omega\le\frac{7-\gamma}{\gamma+1}$.
In order to verify our results we compare them to analytical approximations
and full hydrodynamic simulations. We demonstrate how this method
can be used to describe arbitrary (not just self similar) angular
perturbations.

\noindent This work complements our previous analysis on radial, spherically
symmetric perturbations, and allows one to calculate the response
of an explosion to arbitrary perturbations in the upstream density.
Together, they settle an age old controversy about the inner boundary
conditions.
\end{abstract}

\section{Introduction}

Expanding shock waves are naturally produced by diverse astrophysical
phenomena, such as supernovae, gamma ray bursts, stellar winds, and
more. So far, analytical self similar solutions have been found for
several simple cases, of which we take special interest in the case
of strong spherical shocks propagating into a density profi{}le that
decays as a power of the radius
\begin{equation}
\rho_{a}\left(r\right)=Kr^{-\omega}\label{eq:unpert_density}
\end{equation}
The first solutions of this kind to be found, now commonly known as
the Sedov Taylor Von Neumann solutions \cite{Taylor1950,Sedov1959,Neumann1963},
for the case $\omega<3$ describe decelerating shocks. The solutions
are based on the conservation of energy inside the shocked region,
and they are called type I solutions. If $\omega<\frac{7-\gamma}{\gamma+1}$,
where $\gamma$ is the adiabatic index of the ambient gas, then the
explosion is filled, i.e. the pressure is greater than zero everywhere
inside the shocked region. If $\frac{7-\gamma}{\gamma+1}<\omega<3$,
then the explosion is hollow, i.e. the pressure (and the density)
vanishes at a finite radius \cite{Waxman1993}. If $\omega=\frac{7-\gamma}{\gamma+1}$,
then the hydrodynamic equations admit a relatively simple solution
known as the Primakoff solution \cite{Sedov1977}. 

\noindent The solutions discussed above, while useful, falls short
when describing shocks propagating into density profiles that deviate
from a simple power law decay. This might occur in a variety of astrophysical
scenarios, e.g. supernova shock propagating into a modulated stellar
wind. For this reason it is desirable to generalize as much as possible
the external density profile for which we can obtain analytic solutions,
and this is what we attempt here. This paper takes after a similar
endeavor for type II solutions \cite{Sari2012}, and for radial type
I solutions \cite{Yalinewich2013}. These two cases have to be treated
differently, because of different inner boundary conditions. We note
that while there is a consensus about the inner boundary conditions
in the case of type II explosions \cite{Sari2000}, the inner boundary
conditions in the case of type I explosions have been a bone of contention
for decades \cite{Ryu1991,Bernstein1980,Gaffet1984,Kushnir2005}.

\noindent The plan in this paper is as follows: In §\ref{sec:Density-Perturbations}
we develop the perturbation equations and boundary conditions and
compare the solutions to numerical results from a full hydrodynamic
simuation. In §\ref{sec:Special-Analytical-Results} we present a
few cases where the equations admit an analytic solution. In §\ref{sec:Extension-to-Arbitrary}
we demonstrate how this formalism can be used for any angular perturbation
in the upstream density (not just spherical harmonics). Finally, we
conclude and discuss the results in §\ref{sec:Discussion}.

\section{Density Perturbations\label{sec:Density-Perturbations}}

\subsection{The Perturbation Equations}

For the perturbation equation to be tractable we aim at a self similar
solution by carefully choosing a perturbation whose characteristic
wavelength scales like the radius. Namely, we take the perturbed density
profile to be
\begin{equation}
\rho_{a}\left(r\right)+\delta\rho_{a}\left(r\right)=Kr^{-\omega}\left(1+\varepsilon\left(\frac{r}{r_{0}}\right)^{q}Y_{lm}\left(\theta,\phi\right)\right)\label{eq:ambient_density_pert}
\end{equation}
where $r_{0}$ has dimensions of length and bears only on the phase
of the perturbation, $q$ is the growth rate of the perturbation and
$\varepsilon$ is a small, real and dimensionless amplitude. We take
the real part of any complex quantity to be the physically significant
element.

\noindent We define perturbed flow variables
\begin{equation}
\delta\mathbf{u}\left(r,\theta,\phi,t\right)=\dot{R}\xi\left[\delta U_{r}\left(\xi\right)Y_{lm}\left(\theta,\phi\right)+\delta U_{T}\left(\xi\right)\nabla_{T}Y_{lm}\left(\theta,\phi\right)\right]f\left(t\right)
\end{equation}
\begin{equation}
\delta\rho\left(r,\theta,\phi,t\right)=KR^{-\omega}\delta G\left(\xi\right)Y_{lm}\left(\theta,\phi\right)f\left(t\right)
\end{equation}
\begin{equation}
\delta p\left(r,\theta,\phi,t\right)=KR^{-\omega}\dot{R}^{2}\delta P\left(\xi\right)Y_{lm}\left(\theta,\phi\right)f\left(t\right)
\end{equation}
\begin{equation}
\delta R\left(t\right)=R\left(t\right)Y_{lm}\left(\theta,\phi\right)f\left(t\right)
\end{equation}
Where $\xi=\frac{r}{R}$ is the dimensionless radius, $G\left(\xi\right)$,
$P\left(\xi\right)$ and $U\left(\xi\right)$ are the dimensionless
unperturbed density, pressure and velocity and $\delta G\left(\xi\right)$,
$\delta P\left(\xi\right)$, $\delta U_{r}\left(\xi\right)$, $\delta U_{T}\left(\xi\right)$
are the dimensionless perturbations in the density, pressure, radial
velocity and angular velocity. 

\noindent To allow separation of variables, the function $f\left(t\right)$
must satisfy
\begin{equation}
f\left(t\right)=\frac{\varepsilon}{d}\left(\frac{R}{r_{0}}\right)^{q}\quad\Rightarrow\quad\frac{\dot{f}R}{f\dot{R}}=q
\end{equation}
We note that the boundary conditions at the blast front dictate that
the perturbed density ahead of the shock and the perturbed variables
behind the shock would have the same growth rate $q$. The parameter
$d$ represents the coupling between perturbations in the upstream
to perturbations in the downstream. The larger it is the weaker the
coupling and the downstream perturbation would be weaker. It is determined
by the inner boundary conditions, as described in the next section.

\noindent Plugging the perturbed hydrodynamic variables into the hydrodynamic
equations yields dimensionless ODEs (ordinary differential equations)
for the perturbed variables
\begin{equation}
l\left(l+1\right)\delta U_{T}G-G\left(q-\omega+3U+\xi U'\right)-\xi\left(1-U\right)G\delta U_{r}'+-\delta U_{r}\left(3G+\xi G'\right)=0\label{eq:pert_mass_conservation}
\end{equation}
\begin{equation}
-\delta GP'+\xi\delta U_{r}G^{2}\left(\frac{1}{2}+q-\frac{1}{2}\omega+2U+\xi U'\right)+G\left(\delta P'-\xi^{2}\left(1-U\right)G\delta U_{r}'\right)=0\label{eq:pert_radial_momentum_conservation}
\end{equation}
\begin{equation}
\frac{\delta P}{G}+\xi^{2}\left(\left(\frac{1}{2}+q+2U\right)\delta U_{T}-\xi\left(1-U\right)\delta U_{T}'\right)=0\label{eq:pert_tangential_momentum_conservation}
\end{equation}
\begin{equation}
G\left(\xi\left(\gamma\left(1-U\right)\delta GP'\right)+G\left(\delta U_{r}P'-\left(1-U\right)\delta P'\right)\right)+\label{eq:pert_entropy_conservation}
\end{equation}
\[
\delta P\left(\left(q+3+\omega\gamma\right)G-\gamma\xi\left(-1+U\right)G'\right)-
\]
\[
\gamma P\left(\delta G\left(\left(q+3+\omega\gamma\right)G+\left(\gamma+1\right)\xi\left(1-U\right)G'\right)+\xi G\left(-\left(1-U\right)\delta G'+\delta U_{r}G'\right)\right)=0
\]

\subsection{Boundary Conditions for the Perturbations}

The boundary conditions for the perturbed variables at the blast front
are \cite{Kushnir2005,Ryu1987,Sari2012,Oren2009} 
\begin{equation}
\delta G\left(\xi=1\right)=\frac{\gamma+1}{\gamma-1}\left(d-\omega\right)-G'\left(\xi=1\right)\label{eq:pert_rankine_hugoniot_density}
\end{equation}
\begin{equation}
\delta U_{r}\left(\xi=1\right)=\frac{2}{\gamma+1}q-U'\left(\xi=1\right)\label{pert_rankine_hugoniot_radial_velocity}
\end{equation}
\begin{equation}
\delta U_{t}\left(\xi=1\right)=-\frac{2}{\gamma+1}\label{pert_rankine_hugoniot_tangential_velocity}
\end{equation}
\begin{equation}
\delta P\left(\xi=1\right)=\frac{2}{\gamma+1}\left[2\left(q+1\right)-\omega+d\right]-P'\left(\xi=1\right)\label{pert_rankine_hugoniot_pressure}
\end{equation}
In analogy to the unperturbed solution, where the parameter $\alpha=\frac{d\ln R}{d\ln t}$
(where $R$ is the radius of the shock front and $t$ is the time)
is determined by the inner boundary conditions, the parameter $d$
is also determined by the inner boundary condition. The inner boundary
condition is that the tangential velocity does not diverge there,
and that is achieved only if the pressure perturbation vanishes there
\cite{Ryu1991}.

\noindent If $q$ is imaginary, the real part of $f\left(t\right)$
is periodic, the solution is discretely self similar, i.e. it repeats
itself up to a scaling factor in intervals of $\frac{\Delta R}{R}=\exp\left(\frac{2\pi}{Im\left(q\right)}\right)-1$.
While the unperturbed solution and the perturbations in their complex
form are both self similar, the physical solution which is the real
part of their sum is not.

\subsection{Solution of the Perturbed Equations}

While self similarity simplifies the problem by reducing the PDEs
(partial differential equations) to ODEs, the resulting ODEs, in general,
do not admit analytic solutions. Therefore, for each specific set
of parameters $\gamma$, $\omega$, $l$ and $q$, the functions $\delta G$,
$\delta P$, $\delta U_{r}$, $\delta U_{t}$ and the parameter $d$
are found numerically. Since the ODEs are linear, there exists a matrix
that relates the vector of the values of the flow variables at the
center to the same vector at the front
\begin{equation}
\left(\begin{array}{c}
\delta G\left(\xi=1\right)\\
\delta P\left(\xi=1\right)\\
\delta U_{r}\left(\xi=1\right)\\
\delta U_{t}\left(\xi=1\right)
\end{array}\right)=\mathbf{M}\left(\begin{array}{c}
\delta G\left(\xi=0\right)\\
\delta P\left(\xi=0\right)\\
\delta U_{r}\left(\xi=0\right)\\
\delta U_{t}\left(\xi=0\right)
\end{array}\right)\label{eq:matix_equation}
\end{equation}

\noindent The ODEs are independent of the parameter $d$, or any of
the boundary conditions for that matter. Hence, the matrix $\mathbf{M}$
can be obtained by direct numerical integration of the ODEs.

\noindent We require that the pressure perturbation vanishes at the
center
\begin{equation}
\delta P\left(\xi=0\right)=0\label{eq:pert_centre_bc}
\end{equation}
Thus equations \ref{eq:pert_rankine_hugoniot_density} through \ref{eq:pert_centre_bc}
constitute 5 linear equation for 5 variables ($d$, $\delta G\left(\xi=0\right)$,
$\delta P\left(\xi=0\right)$, $\delta U_{r}\left(\xi=0\right)$ and
$\delta U_{t}\left(\xi=0\right)$). Solving these equations yield
the value of $d$.

\begin{figure}
\begin{centering}
\includegraphics[width=13cm,height=13cm,keepaspectratio]{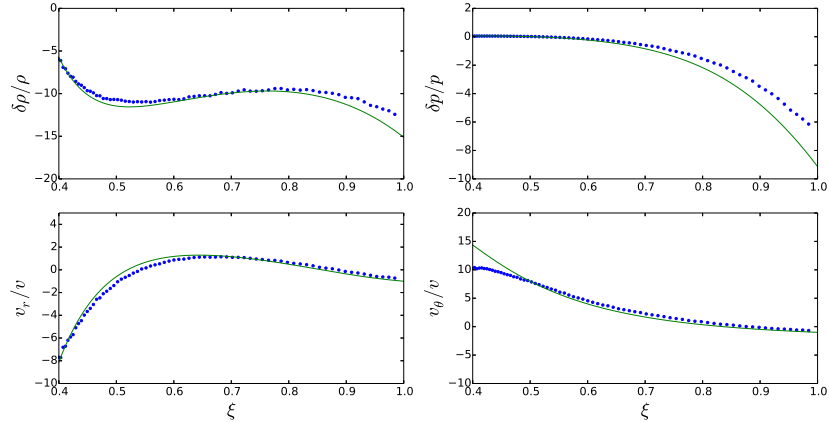}
\par\end{centering}

\caption{Comparison of the analytic (green solid line) and numeric profiles
(blue dots) of the perturbed hydrodynamic variables: density (top
left), pressure (top right), radial velocity (bottom left) and angular
velocity (bottom right). The horizontal axis is the dimensionless
radius (radius divided by the shock radius). The explosion parameters
are $\gamma=\frac{5}{3}$, $\omega=0$, $q=0$, $l=1$, $\varepsilon=0.1$.
In order to extract the information from the two dimensional simulations,
we fit each ring to an expression of the form $A_{0}+A_{1}\cos\theta$
and plotted the ratio $\frac{A_{1}}{A_{0}}$ versus radius (except
for the tangential velocity where we used $\sin\theta$ instead of
$\cos\theta$, and normalized by the coefficient $A_{0}$ of the radial
velocity). The inconsistency near the shock front stems from the fact
that some of the ring is in the upstream. \label{fig:numeric_vs_analytic}}
\end{figure}

\noindent A comparison between the solutions discussed above and a
hydrodynamic simulation is presented in figure \ref{fig:numeric_vs_analytic},
for a perturbation with the following parameters $\gamma=\frac{5}{3}$,
$\omega=0$, $q=0$, $l=1$ and $\varepsilon=0.1$ . All curves seem
to agree. The numerical calculations was carried out using the hydrocode
RICH {[}Yalinewich, Steinberg \& Sari, in perperation{]}. The initial
grid consisted of points at fixed angular intervals $\Delta\theta=\frac{2\pi}{100}$
along a logarithmic, i.e. the radius of the $n$th cell is given by
$r_{n}=r_{0}e^{n\Delta\theta}$ where $r_{0}=10^{-3}$ and points
outside the computational domain $\left(x,y\right)\in\left[0,2\right]\times\left[-2,2\right]$
were omitted. In order to extract the perturbation from the 2D numerical
data, we projected the raw unstructured data into a series of concentric
rings, and fit the values of each ring to an expression of the form
$\underset{l}{\sum}A_{l}Y_{l0}\left(\theta,0\right)$ and the result
of the fit are the coefficients $A_{l}$. In our case, $l=1$ so the
expression we fit to is $A_{0}+A_{1}\cos\theta$, and the perturbation
is given by the ratio $\frac{A_{1}}{A_{0}}$, except for the tangential
velocity, where the expression was $A_{1}\sin\theta$ and for normalization
use the coefficient $A_{0}$ of the radial velocity.

\noindent Figure \ref{fig:numeric_vs_analytic} shows that the wavelength
of the density fluctuations is shorter than those of the pressure
and velocity. This happens because the density is affected by both
traveling sound waves and entropy waves, while the pressure and velocity
are affected solely by sound waves. From this argument it follows
that the characteristic wavelength are given by $\frac{2\pi}{\Im\left(q\right)}\left(1-\xi U\pm\sqrt{\gamma\frac{P}{G}}\right)$
for the pressure and velocity, together with $\frac{2\pi}{\Im\left(q\right)}\left(1-\xi U\right)$
for density perturbations.

\noindent 

\section{Analytical Results for Special Cases\label{sec:Special-Analytical-Results}}

Though for a general choice of the parameters $\omega$, $q$, $l$
and $\gamma$ a numerical method must be employed to determine $d$,
in some special cases it is possible to obtain an explicit analytic
expression for $d$. In this section we will present few such cases
which we were able to find.

\subsection{Shifted Explosion}

We consider a spherically symmetric explosion in a coordinate system
where the origin is offset by $\varepsilon r_{0}\hat{z}$ to the hot
spot. In such a coordinate system the ambient density profile, to
first order in $\varepsilon$ 
\begin{equation}
\rho_{i}\left(r\right)\approx kr^{-\omega}\left(1+\omega\varepsilon\frac{r_{0}}{r}\cos\theta\right)
\end{equation}
where $\cos\theta=\hat{r}\cdot\hat{z}$. The radius of the shock front,
to first order in $\varepsilon$, is given accordingly by
\begin{equation}
R+\delta R=R\left(1+\varepsilon\cos\theta\right)
\end{equation}
so
\begin{equation}
d\left(l=1,q=-1\right)=\omega\label{eq:d_shifted_explosion}
\end{equation}
The hydrodynamic variables can be obtained from the unperturbed solutions
in a similar manner
\begin{equation}
\delta P\left(\xi\right)=-P'\left(\xi\right)
\end{equation}
\begin{equation}
\delta G\left(\xi\right)=-G'\left(\xi\right)
\end{equation}
\begin{equation}
\delta U_{r}\left(\xi\right)=-\frac{U\left(\xi\right)}{\xi}-U'\left(\xi\right)
\end{equation}
\begin{equation}
\delta U_{t}\left(\xi\right)=-\frac{U\left(\xi\right)}{\xi}
\end{equation}

\noindent It is easy to verify that these expressions satisfy the
differential equations \ref{eq:pert_mass_conservation}-\ref{eq:pert_entropy_conservation}.
This result supports the idea that at the center $\delta P=0$. The
reason for that is that in the case of a filled type I explosion,
the pressure plateaus as the radius approaches zero, so $\underset{\xi\rightarrow0}{\lim}P'\left(\xi\right)=0$
and hence $\delta P\left(0\right)=0$.

\subsection{Thin Shell Model}

In the limit $\gamma\rightarrow1$ it is possible to use the thin
shell approximation to find an analytic relation for $d$. Following
\cite{Ryu1987}, we define 
\begin{equation}
\delta=\frac{\sigma-\sigma_{0}}{\sigma_{0}}
\end{equation}
\begin{equation}
\Delta R=R-R_{0}
\end{equation}
where
\begin{equation}
\sigma_{0}=\frac{1}{R^{2}}\int_{0}^{R}\rho r^{2}dr
\end{equation}
\begin{equation}
\sigma=\frac{1}{\left(R+\delta R\right)^{2}}\int_{0}^{R+\delta R}\left(\rho+\delta\rho\right)r^{2}dr
\end{equation}
are the unperturbed and perturbed surface density. The perturbation
equations are
\begin{equation}
\frac{\partial\delta}{\partial t}=-\frac{2}{R_{0}}\frac{\partial\Delta R}{\partial t}+2\frac{\dot{R}_{0}}{R_{0}^{2}}\Delta R+\frac{\Delta\rho}{\sigma_{0}}\dot{R}_{0}-\omega\frac{\rho_{0}}{\sigma_{0}}\frac{\Delta R}{R_{0}}\dot{R}_{0}+\frac{\rho_{0}}{\sigma_{0}}\frac{\partial\Delta R}{\partial t}-\nabla_{T}\mathbf{v}_{T}-\delta\frac{\rho_{0}}{\sigma_{0}}\dot{R}_{0}
\end{equation}
\begin{equation}
\frac{\partial^{2}\Delta R}{\partial t^{2}}=-\delta\dot{R}_{0}-2\frac{\rho_{0}\dot{R}_{0}}{\sigma_{0}}\frac{\partial\Delta R}{\partial t}-\frac{\Delta\rho\dot{R}_{0}^{2}}{\sigma_{0}}+\omega\frac{\rho_{0}}{\sigma_{0}}\frac{\Delta R}{R_{0}}\dot{R}_{0}^{2}
\end{equation}
\begin{equation}
\frac{\partial\mathbf{v}_{T}}{\partial t}=-\frac{\rho_{0}\dot{R}_{0}}{\sigma_{0}}\mathbf{v}_{T}-\frac{\dot{R}_{0}}{R_{0}}\mathbf{v}_{T}-\frac{1}{R_{0}}\frac{P_{i}}{\sigma_{0}}\nabla_{T}\Delta R
\end{equation}
where $\Delta\rho=\rho-\rho_{0}$. Assuming $\Delta\rho,\Delta R,\delta\propto Y_{lm}\left(\theta,\phi\right)R_{0}^{q}$
and $\mathbf{v}_{T}\propto\nabla_{T}Y_{lm}\left(\theta,\phi\right)R_{0}^{q}$
we can solve for the coefficients and find the parameter $d$ using
the ratio $\Delta\rho/\Delta R$
\begin{equation}
d\left(\gamma=1\right)=(s+1)\left(\omega^{2}-8\omega+15\right)(s(\omega-5)+\omega-3)/
\end{equation}
\[
\left[4l\left(l+1\right)(\omega-3)^{2}+(s+1)(\omega-5)\left(s^{3}(\omega-5)^{3}+\right.\right.
\]
\[
\left.\left.s^{2}(5\omega-17)(\omega-5)^{2}+s\left(8\omega^{3}-90\omega^{2}+328\omega-390\right)+4(\omega-3)^{2}(\omega-2)\right)\right]
\]
where $s=q\alpha$ and $\alpha=\frac{2}{5-\omega}$. This relation
reproduces equation \ref{eq:d_shifted_explosion} for $l=1$ and $q=-1$.

\subsection{Primakoff Solution}

\begin{figure}
\begin{centering}
\includegraphics[width=9cm,height=9cm,keepaspectratio]{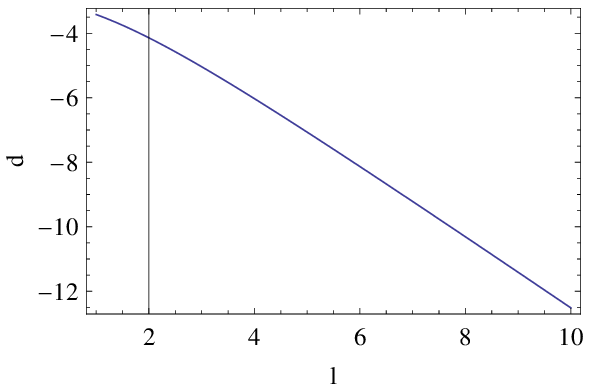}
\par\end{centering}

\begin{centering}
\includegraphics[width=9cm,height=9cm,keepaspectratio]{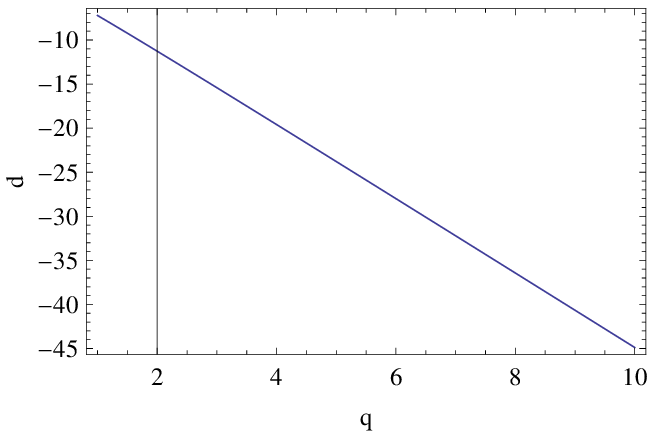}
\par\end{centering}

\caption{$d$ as a function of $l$ for $\gamma=\frac{5}{3}$ and $q=0$ (top)
and $d$ as a function of $q$ for $\gamma=\frac{5}{3}$ (bottom),
both for the Primakoff explosion\label{fig:prim_d_vs_q}}
\end{figure}

In the case of the Primakoff explosion, the perturbation equations
can be solved analytically. With the substitution
\begin{equation}
\mathbf{Y}=\left(\frac{\delta G}{G},\frac{\delta P}{P},\frac{\delta U_{r}}{U},\frac{\delta U_{T}}{U}\right)^{T}
\end{equation}
the system of ODEs can be reduced to the form
\begin{equation}
\frac{d\mathbf{Y}}{d\ln\xi}=\mathbf{M}\cdot\mathbf{Y}
\end{equation}
\begin{equation}
\mathbf{M}=\left(\begin{array}{cccc}
\frac{q(\gamma+1)^{2}+6(\gamma-1)}{\gamma^{2}-1} & -\frac{2(\gamma q+q+3\gamma-3)}{\gamma^{2}-1} & -\frac{2(\gamma q+q-\gamma+7)}{\gamma^{2}-1} & \frac{2l(l+1)}{\gamma+1}\\
\frac{6\gamma}{\gamma+1} & -\frac{\gamma q+q+6\gamma}{\gamma+1} & -\frac{2\left(q\gamma^{2}+(q+5)\gamma-3\right)}{\gamma^{2}-1} & \frac{2l(l+1)\gamma}{\gamma+1}\\
\frac{3(\gamma-1)}{\gamma+1} & -\frac{\gamma q+q+3\gamma-3}{\gamma+1} & -\frac{\gamma q+q+3\gamma+11}{\gamma+1} & \frac{2l(l+1)\gamma}{\gamma+1}\\
0 & 1 & 0 & \frac{\gamma q+q-3\gamma+5}{\gamma-1}
\end{array}\right)\label{eq:primakoff_derivative_matrix}
\end{equation}
The general solution is
\begin{equation}
\mathbf{Y}\left(\xi\right)=\exp\left(\mathbf{M}\ln\xi\right)\mathbf{Y}\left(1\right)
\end{equation}
Every term in $\mathbf{Y}\left(\xi\right)$ is the sum of 4 power
laws in $\xi$, and the powers are eigenvalues. The value at the shock
front is determined by the Rankine Hugoniot conditions (equations
\ref{eq:pert_rankine_hugoniot_density},\ref{pert_rankine_hugoniot_radial_velocity},\ref{pert_rankine_hugoniot_tangential_velocity},\ref{pert_rankine_hugoniot_pressure})
and the inner boundary conditions that the pressure perturbation vanishes.
Usually, out of the 4 eigenvector modes, one would diverge at the
center (we denote the well behaved modes by $\mathbf{Y}_{1}$, $\mathbf{Y}_{2}$
and $\mathbf{Y}_{3}$, and the diverging mode by $\mathbf{Y}_{4}$).
To prevent the divergence, we require that the solution will be a
linear superposition of only the well behaved modes 
\begin{equation}
\mathbf{Y}\left(1\right)=\sum_{i=1}^{3}a_{i}\mathbf{Y}_{i}
\end{equation}
This gives us 4 linear equations with 4 variables ($a_{1}$,$a_{2}$,
$a_{3}$ and $d$), from which we can extract the value of the paramter
$d$. Unfortunately, the expression for the parameter $d$ is too
long to be written here. We evaluate $d$ numerically as a function
of both $q$ and $l$ for $\gamma=\frac{5}{3}$ and show the results
in figure \ref{fig:prim_d_vs_q}. If we interpret $q$ as the radial
wave number, and $l$ as the angular wave number, then from figure
\ref{fig:prim_d_vs_q} it seems that the magnitude of $d$ increases
linearly with $q$ and $l$. Using the appropriate approximations
for large $l$ reveals that in that limit $\underset{l\rightarrow\infty}{\lim}\frac{d}{l}=-\frac{\sqrt{2\gamma}}{\sqrt{\gamma+1}}$.
We note that in the case of the Primakoff explosion the speed of sound
vanishes at the center, whereas in the general filled type I explosion
the speed of sound diverges there, so in the Primakoff explosion there
are no reflections from the center, while the general case has them.
For that reason, the general explosion does not have this asymptotic
behavior.

\noindent 

\section{Extension to Arbitrary Angular Dependence\label{sec:Extension-to-Arbitrary}}

\begin{figure}
\noindent \begin{centering}
\includegraphics[width=9cm,height=9cm,keepaspectratio]{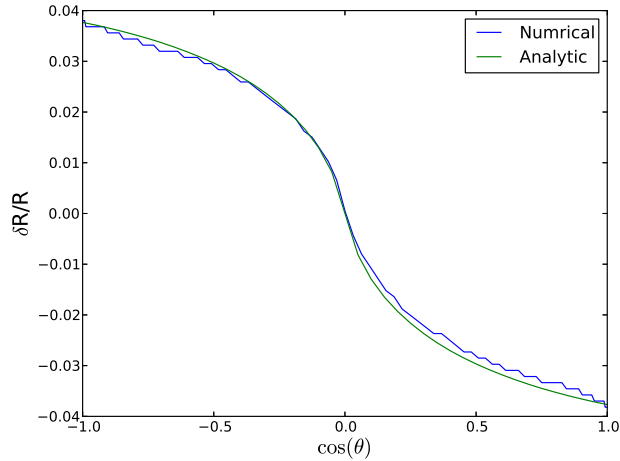}
\par\end{centering}

\caption{A comparison between the angular profiles of the analytic perturbation
to the shock front (green) and the numerical simulation (blue) in
case of an explosion between two uniform density half spaces with
small density difference between them ($\sigma=\frac{\rho_{2}-\rho_{1}}{\rho_{1}}=0.1$).
\label{fig:crater}}
\end{figure}

The formalism presented so far is limited to just one angular mode.
However, due to linearity, any perturbation can be decomposed into
spherical harmonics and each mode solved for individually. We demonstrate
this using a problem similar to that used in the case of density perturbation
to type II explosion \cite{Sari2012}. The problem we are considering
is an explosion that happens on the planar interface between two half
spaces. Each half space has uniform density, but there is a slight
difference between the densities of each of the half spaces. The ambient
density profile can thus be described by the formula $\rho_{a}\left(r,\theta\right)=\rho_{0}\left(1+\sigma\Theta\left(\theta\right)\right)$,
where $\Theta\left(x\right)$ is the Heaviside step function and $\sigma$
and $\rho_{0}$ are constants. Such expression can be expanded in
spherical harmonics
\begin{equation}
\Theta\left(\theta\right)=\sum_{n=0}^{\infty}\frac{\pi\sqrt{4n+3}}{\Gamma\left(\frac{1}{2}-n\right)\Gamma\left(2+n\right)}Y_{2n+1,0}\left(\theta,0\right)
\end{equation}
The shape of the perturbation to the shock front is given by
\begin{equation}
\frac{\delta R\left(\theta,t\right)}{R\left(t\right)}=\sigma\sum_{n=0}^{\infty}\frac{1}{d\left(2n+1\right)}\frac{\pi\sqrt{4n+3}}{\Gamma\left(\frac{1}{2}-n\right)\Gamma\left(2+n\right)}Y_{2n+1,0}\left(\theta,0\right)\label{eq:analytic_crater}
\end{equation}
 In order to verify this result, we ran a numerical simulation for
this scenario with $\sigma=0.1$ and compared it to the analytic result
(equation \ref{eq:analytic_crater}) calculated up to order $n=100$.
The results are plotted in figure \ref{fig:crater}, and there seems
to be a good fit between the two methods, and they get closer as the
resolution of the numerical simulation increases (the simulation we
did had 1000 cells in the radial direction and 100 cells in the angular
direction).

\noindent 

\section{Discussion\label{sec:Discussion}}

We have laid out a method for solving the strong explosion problem
in density profiles that deviate from a pure radial power law dependence.
The key lies in choosing radially log-periodic perturbations which
do not introduce a new scale into the problem. This leads to self
similar perturbation in the hydrodynamic quantities behind the shock,
which can be found by solving a set of ordinary differential equations.
The perturbations are fully self similar when the density perturbation
is given in equation \ref{eq:ambient_density_pert}, but if $q$ is
imaginary, then the solution is only discretely self similar because
of the periodic nature of the perturbations.

\noindent The inner boundary condition in radial perturbation differs
from that proposed here. We recall that in the case of radial perturbations
to filled type I explosions the inner boundary condition is $\delta U_{r}\left(\xi=0\right)=0$,
as given from the requirement that the total energy remains constant.
This condition cannot be used for angular perturbation, because the
total contribution of the perturbations to the energy is always zero
for $l=0$. From the other end, the condition that the tangential
velocity does not diverge cannot be applied to radial perturbations,
as the tangential velocity is always zero.

\noindent The linearized perturbation treatment naturally ensures
that the perturbations will be linear in $\varepsilon$ (and will
contain no higher power of $\varepsilon$). This simplifies the solution
of the problem but limits the validity of the method to small perturbations.
The perturbation theory developed above fails when $\varepsilon$
becomes too large. The deviation from linear theory is of order $\varepsilon^{2}$.


\begin{thebibliography}{10}

\bibitem{Bernstein1980}
I.~B. {Bernstein} and D.~L. {Book}.
\newblock {Stability of the Primakoff-Sedov blast wave and its
  generalizations}.
\newblock {\em \apj}, 240:223--234, August 1980.

\bibitem{Gaffet1984}
B.~{Gaffet}.
\newblock {Stability of Self-Similar Flow - Correct Form of the Basic Equations
  and of the Shock Boundary Conditions}.
\newblock {\em \apj}, 279:419, April 1984.

\bibitem{Kushnir2005}
D.~{Kushnir}, E.~{Waxman}, and D.~{Shvarts}.
\newblock {The Stability of Decelerating Shocks Revisited}.
\newblock {\em \apj}, 634:407--418, November 2005.

\bibitem{Landau1959}
L.~D. {Landau} and E.~M. {Lifshitz}.
\newblock {\em {Fluid mechanics}}.
\newblock 1959.

\bibitem{Mignone2007}
A.~{Mignone}, G.~{Bodo}, S.~{Massaglia}, T.~{Matsakos}, O.~{Tesileanu},
  C.~{Zanni}, and A.~{Ferrari}.
\newblock {PLUTO: A Numerical Code for Computational Astrophysics}.
\newblock {\em apjs}, 170:228--242, May 2007.

\bibitem{Neumann1963}
John~Von Neumann, A.~W. Taub, and A.~H. Taub.
\newblock {\em The Collected Works of John Von Neumann: 6-Volume Set}.
\newblock Reader's Digest Young Families, 1963.

\bibitem{Oren2009}
Y.~{Oren} and R.~{Sari}.
\newblock {Discrete self-similarity in type-II strong explosions}.
\newblock {\em Physics of Fluids}, 21(5):056101--+, May 2009.

\bibitem{Ryu1987}
D.~{Ryu} and E.~T. {Vishniac}.
\newblock {The growth of linear perturbations of adiabatic shock waves}.
\newblock {\em apj}, 313:820--841, February 1987.

\bibitem{Ryu1991}
D.~{Ryu} and E.~T. {Vishniac}.
\newblock {The dynamic instability of adiabatic blast waves}.
\newblock {\em apj}, 368:411--425, February 1991.

\bibitem{Sari2012}
R.~{Sari}, N.~{Bode}, A.~{Yalinewich}, and A.~{MacFadyen}.
\newblock {Slightly two- or three-dimensional self-similar solutions}.
\newblock {\em Physics of Fluids}, 24(8):087102, August 2012.

\bibitem{Sari2000}
R.~{Sari}, E.~{Waxman}, and D.~{Shvarts}.
\newblock {Shock Wave Stability in Steep Density Gradients}.
\newblock {\em \apjs}, 127:475--479, April 2000.

\bibitem{Sedov1959}
L.~I. {Sedov}.
\newblock {\em {Similarity and Dimensional Methods in Mechanics}}.
\newblock 1959.

\bibitem{Sedov1977}
L.~I. {Sedov}.
\newblock {Similarity methods and dimensional analysis in mechanics /8th
  revised edition/}.
\newblock {\em Moscow Izdatel Nauka}, 1977.

\bibitem{Taylor1950}
Geoffrey Taylor.
\newblock The formation of a blast wave by a very intense explosion. i.
  theoretical discussion.
\newblock {\em Proceedings of the Royal Society of London. Series A.
  Mathematical and Physical Sciences}, 201(1065):159--174, 1950.

\bibitem{Waxman1993}
E.~{Waxman} and D.~{Shvarts}.
\newblock {Second-type self-similar solutions to the strong explosion problem}.
\newblock {\em Physics of Fluids}, 5:1035--1046, April 1993.

\bibitem{Yalinewich2013}
Almog Yalinewich and Reem Sari.
\newblock Discrete self similarity in filled type i strong explosions.
\newblock {\em Physics of Fluids (1994-present)}, 25(12):--, 2013.

\end{thebibliography}
\end{document}